\documentclass[journal=jacsat,manuscript=article]{achemso}
\usepackage{chemformula} 
\usepackage[T1]{fontenc} 

\author{Guilherme S. L. Fabris}
\affiliation{Applied Physics Department and Center for Computational Engineering \& Sciences, State University of Campinas, Campinas, São Paulo 13083970, Brazil.}

\author{Raphael B. de Oliveira}
\affiliation{Department of Materials Science and NanoEngineering, Rice University, Houston, TX 77005, USA.}

\author{Marcelo L. Pereira Jr}
\affiliation{Department of Materials Science and NanoEngineering, Rice University, Houston, TX 77005, USA.}
\alsoaffiliation{University of Bras\'{i}lia, College of Technology, Department of Electrical Engineering, Bras\'{i}lia, Federal District, Brazil.}
\email{mp200@rice.edu}

\author{Robert Vajtai}
\affiliation{Department of Materials Science and NanoEngineering, Rice University, Houston, TX 77005, USA.}

\author{Pulickel M. Ajayan}
\affiliation{Department of Materials Science and NanoEngineering, Rice University, Houston, TX 77005, USA.}

\author{Douglas S. Galvão}
\affiliation{Applied Physics Department and Center for Computational Engineering \& Sciences, State University of Campinas, Campinas, São Paulo 13083970, Brazil.}
\alsoaffiliation{Center for Computational Engineering \& Sciences (CCES), State University of Campinas, Campinas, São Paulo 13083970, Brazil.}
\email{galvao@ifi.unicamp.br}

\title{
From Glaphene to Glaphynes: A Hybridization of 2D Silica Glass and Graphynes
}

\begin{document}

\begin{tocentry}

\centering\includegraphics[width=0.60\linewidth]{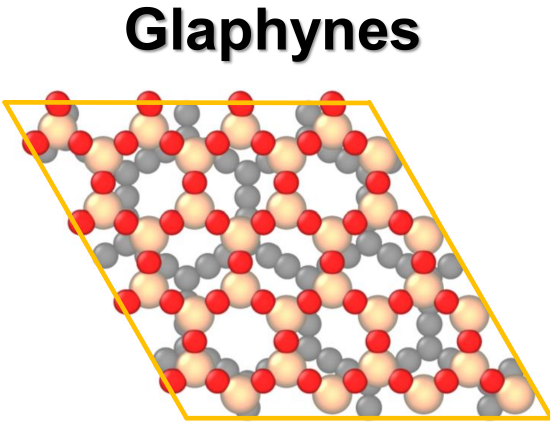}

New Materials: Glaphynes, from Glaphene.

\end{tocentry}

\begin{abstract}
Hybrid two-dimensional (2D) materials have attracted increasing interest as platforms for tailoring electronic properties through interfacial design. Very recently, a novel hybrid 2D material termed glaphene, which combines monolayers of 2D silica glass and graphene, was experimentally realized. Inspired by glaphenes, we proposed a new class of similar structures named glaphynes, which are formed by stacking SiO$_2$ monolayers onto $\alpha$-, $\beta$-, and $\gamma$-graphynes. Graphynes are 2D carbon allotropes with the presence of acetylenic groups (triple bonds). The glaphynes' structural and electronic properties were investigated using the density functional tight-binding (DFTB) method, as implemented in the DFTB+ package. Our analysis confirms their energetic and structural stability. We have observed that in the case of glaphynes, the electronic proximity effect can indeed open the electronic band gap, but not for all cases, even with the formation of Si-O-C bonds between silica and graphynes. \end{abstract}

\section{Introduction}

Few fields have experienced such rapid growth in recent decades as nanoscience. This growth can be partly attributed to the prediction and subsequent synthesis of two-dimensional (2D) nanostructures, such as graphene \cite{novoselov2004electric}. Owing to its remarkable properties, including high mechanical strength \cite{lee2008measurement} and thermal conductivity \cite{balandin2011thermal}, graphene has inspired the search for new 2D materials with similarly outstanding characteristics. In this perspective, graphynes \cite{baughman1987structure,LI2020619, C2NR31644G} represent a class of 2D carbon allotropes composed of sp and sp$^2$ hybridized carbon atoms, resulting from the insertion of acetylenic linkers between the carbon atoms of graphene \cite{graphyne/acsami.8b03338}. The proportion and arrangement of these acetylenic chains give rise to distinct types of graphyne, such as $\alpha$, $\beta$, and $\gamma$-graphyne \cite{baughman1987structure,PUIGDOLLERS2016879}, among others. Since the experimental synthesis of $\gamma$-graphyne \cite{li2010rsc,li2018synthesis,RodionovJACS2022,RayPNAS2025}, several studies have proposed its use in supercapacitors \cite{yang2019mechanochemical}, batteries \cite{ding2020ultrasound}, photovoltaic devices \cite{GraphyneElectronics, GraphyneElectronic}, and other potential applications. Further investigations have confirmed the semiconducting nature of $\gamma$-graphyne, with a moderate band gap \cite{narita1998optimized, ALI2024117454,ma11020188,RodionovJACS2022}. Beyond carbon-based systems, other families of 2D nanomaterials have emerged, such as silicates, with special focus on silica (SiO$_2$) \cite{SiO,ultrathinSilica}. Silicon-based nanostructures have played a central role in the development of modern electronics, and silicon oxides are characterized by robust insulating behavior and a wide band gap of approximately 6.7 eV \cite{SiOSiO2, SiO}.

The use of heterostructures to enhance the properties of nanomaterials has increased significantly in recent years. Several studies have proposed the use of different nanostructures to achieve tailored functionalities. For instance, the insertion of hexagonal boron nitride (h-BN) domains into graphene sheets \cite{cbn3, MIRANDA2021114174,ci2010atomic} enables control over mechanical properties \cite{SINGH2023110001, D1RA05831B}. Other examples include graphene/graphyne heterojunctions designed for carbon-based transistors \cite{Graphene/Graphyne,huang2022two}, and heterojunctions between different types of graphynes to tune thermal conductivity \cite{graphyne/Graphyne}. Another widely used approach for creating heterostructures is the layer-by-layer stacking of monolayers. For example, Sun \textit{et al.} \cite{sun2015vertical} investigated graphyne stacked on XSe$_2$ ($\text{X}= \text{Mo}$, W) to engineer the electronic band gap. In contrast, Bhattacharya \textit{et al.} \cite{BHATTACHARYA201673} proposed graphyne-graphene nitride heterostructures for nanocapacitor applications. Independently, Huang \cite{huang2012direct}, Wang \cite{wang2015novel}, and their respective collaborators investigated low-dimensional silica interfaces and their interactions with graphene. Overall, the combination of nanostructures via heterojunctions, embedded domains, or vertical stacking has proven to be an effective strategy for producing materials with intermediate or enhanced properties compared to their parent components.

Very recently, Iyengar \textit{et al.} \cite{Iyengar2025} introduced a novel hybrid 2D material termed glaphene, which combines monolayers of 2D silica glass and graphene. Initially proposed through first-principles calculations and later synthesized via a scalable vapor-phase growth method, glaphene represents a significant experimental milestone in the development of mixed-component 2D materials. Notably, the interlayer interactions in glaphene surpass typical van der Waals forces, leading to strong interlayer hybridization. This hybridization induces a pronounced modification in the electronic structure. While graphene is a semimetal with a zero band gap and 2D silica glass is an insulator with a band gap of approximately 8.2~eV \cite{Iyengar2025}, their combination results in a semiconducting material with a sizable band gap of about 3.6~eV, primarily governed by out-of-plane p$_{z}$ orbital interactions. They demonstrated that it is possible to create electronic band engineering through electronic proximity effects.

In this work, we propose and characterize a new family of heterostructures inspired by the recent proposition and synthesis of glaphene. These new heterostructures are created by stacking a SiO$_2$ monolayer onto a graphyne monolayer. We tested three distinct graphyne configurations, referred to as $\alpha$-, $\beta$-, and $\gamma$-glaphyne, see Fig. \ref{fig:fig1_glaphynes_structure}. We then investigated their energetic and structural stability in order to evaluate their potential experimental feasibility. In addition, we have analyzed their electronic properties, showing that the stacking approach enables the emergence of semiconductors with appreciable electronic band gaps, distinct from those of the individual constituents.

\section{Methodology}

Computational simulations were carried out using the self-consistent-charge (SCC) density-functional tight-binding (DFTB) approximation \cite{Elstner1998}, as implemented in the DFTB+ code \cite{dftb2020}. DFTB+ \cite{Elstner1998,manzano2012, Kubar2013} allows quantum simulations of electronic and structural properties for relatively large systems, offering computational efficiency comparable to traditional tight-binding approaches while achieving accuracy similar to Density Functional Theory (DFT) in specific cases \cite{manzano2012,dftb_performance}. The methodology relies on a second-order expansion of the Kohn-Sham total energy, as extensively discussed in the literature \cite{koskinen2009beginners,dftb2020}. In this study, we employed the matsci-0-3 parametrization \cite{frenzel2004semi}, developed for materials science applications, including graphynes and silicon dioxide nanostructures. Atomic structure visualizations were generated using OVITO \cite{Stukowski2009}, and orbital representations were obtained with the VESTA code \cite{vesta}.

Geometry optimizations were performed using the conjugate gradient algorithm implemented in DFTB+, with convergence criteria set to $10^{-8}$ a.u. for SCC interactions and $10^{-5}$ a.u. for atomic forces. The first Brillouin zone was sampled using a $10\times10\times1$ Monkhorst-Pack grid \cite{monkhorst_pack}, where the sixfold sampling was applied along the periodic direction of the structures. Dispersion interactions were included through the Lennard-Jones potential \cite{Zhechkov2005}, combined with the Universal Force Field (UFF) parametrization \cite{Rappe1992}. Isosurface plots were generated using an isovalue resolution of 0.001~\AA$^{-3}$.

\begin{figure}[t!]
\centering
\includegraphics[width=0.4\linewidth, keepaspectratio]{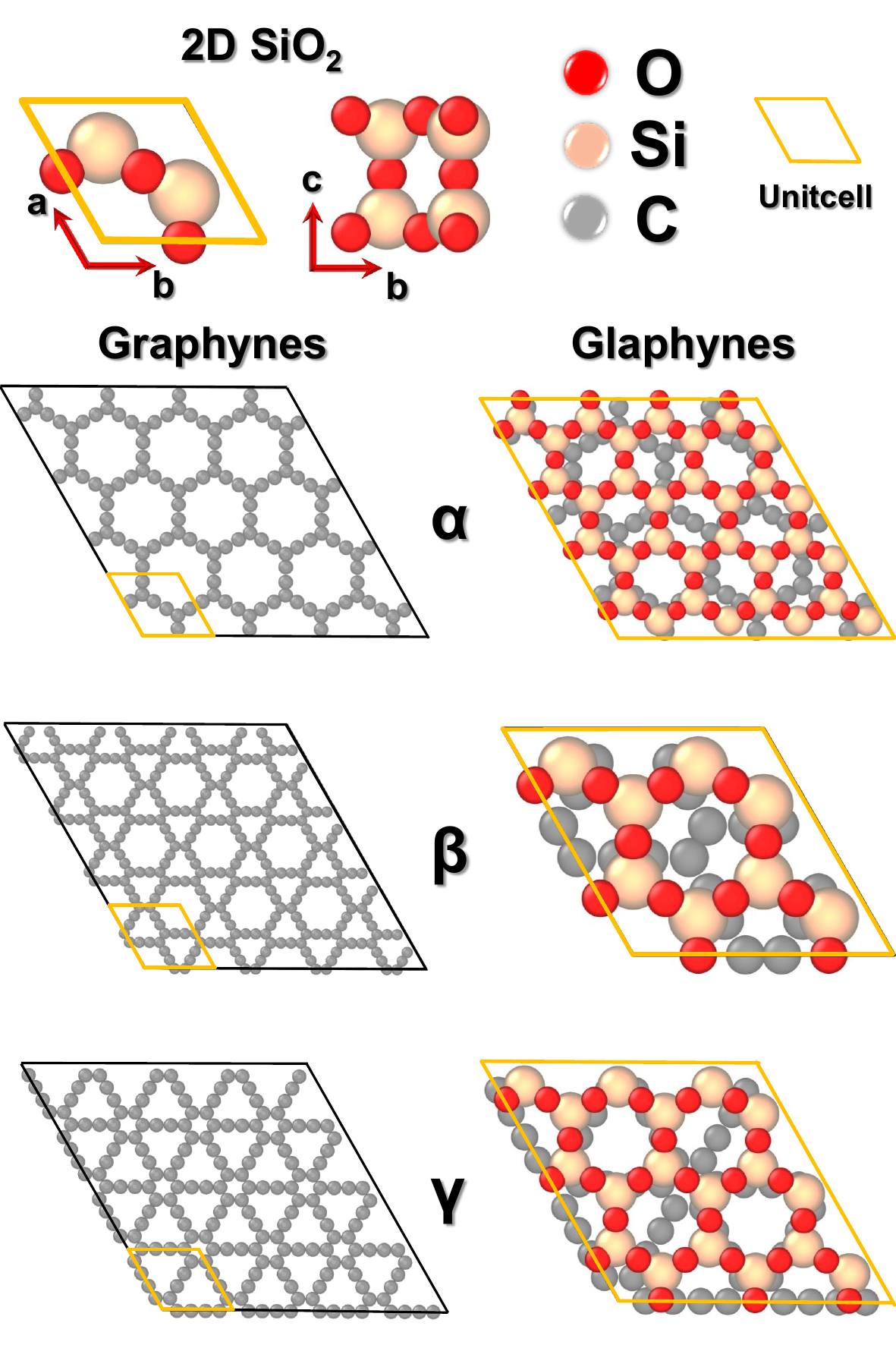}
\caption{Structural representation of two-dimensional SiO$_2$, including $\alpha$-, $\beta$-, and $\gamma$-graphynes (glaphynes). The graphynes are depicted using their respective supercells, while the glaphynes are illustrated through their unit cells, which are highlighted by the orange hexagonal outline.}
\label{fig:fig1_glaphynes_structure}
\end{figure}

The creation of glaphyne heterostructures began with computational modeling of 2D silica, and the $\alpha$-, $\beta$-, and $\gamma$-graphyne structures. This preliminary stage aimed to extract key physical properties and obtain optimized geometries in agreement with reference data. Subsequently, supercells were generated for each graphyne type and the silica structure. The supercell sizes were selected in order to minimize the lattice mismatch between graphyne and silica, thus minimizing strain accumulation at the periodic boundaries. Figure~\ref{fig:fig1_glaphynes_structure} illustrates the created graphyne supercells (in gray), the 2D silica unit cell (in orange), and the resulting glaphyne configurations.

\section{Results and Discussion}

Initially, we performed separate geometry optimizations for the isolated monolayers of SiO$_2$, and the three graphyne types ($\alpha$, $\beta$, and $\gamma$) in order to validate the computational setup. The resulting structural parameters were then compared with the available literature data, as summarized in Table~\ref{table:structural_stability}. The results indicate that the chosen parametrization yields good accuracy, with lattice parameter deviations of approximately 2.5\%, 0.57\%, 0.74\%, and 0.73\% for SiO$_2$, $\alpha$-, $\beta$-, and $\gamma$-graphyne, respectively. For the $\gamma$ angle, an almost negligible deviation was observed across all structures. The calculated bond lengths along the acetylenic chains were 1.411, 1.234, and 1.411~\r{A} for $\alpha$-graphyne; 1.427, 1.227, and 1.427~\r{A} for $\beta$-graphyne; and 1.438, 1.221, and 1.438~\r{A} for $\gamma$-graphyne. For the SiO$_2$ monolayer, the average Si-O bond length was found to be 1.63~\r{A}, in close agreement with experimental data \cite{Zhong2022}. The thickness of the silica layer was estimated to be 4.33~\r{A}, which is consistent with the experimental value reported by Iyengar \textit{et al.} \cite{Iyengar2025}. These consistent results validate the reliability of our computational methodology in describing the structural features of both carbon- and silicon-based two-dimensional systems.

Following the preliminary optimizations, the creation of the glaphynes was carried out by assembling a $3 \times 3$ supercell of $\alpha$- and $\gamma$-graphynes, and a unit cell of $\beta$-graphyne. These were combined with a $4 \times 4$ supercell of 2D SiO$_2$ for $\alpha$- and $\gamma$-graphynes, and a $2 \times 2$ supercell for $\beta$-graphyne. The SiO$_2$ layers were placed above the graphyne structures at an initial vertical separation of approximately 3.5~\r{A}. After geometry optimization, the resulting glaphynes exhibited P6MM symmetry, and only slight modifications were observed in the acetylenic chain bond lengths. The optimized bond lengths for $\alpha$-glaphyne were 1.421, 1.239, and 1.421~\r{A}; for $\beta$-glaphyne, 1.576, 1.247, and 1.576~\r{A}; and for $\gamma$-glaphyne, 1.457, 1.228, and 1.457~\r{A}. The resulting vertical distances between the SiO$_2$ and graphyne layers were 3.16, 3.22, and 3.12~\r{A} for $\alpha$-, $\beta$-, and $\gamma$-glaphynes, respectively. Additionally, the thickness of the SiO$_2$ layer after deposition was found to be approximately 4.39~\r{A}, in close agreement with the experimental value of 0.43~nm reported by Iyengar \textit{et al.}~\cite{Iyengar2025}. Interestingly, the mismatch in the $\beta$-glaphyne case can be significantly reduced by adopting a larger configuration involving a $7 \times 7$ supercell of 2D SiO$_2$ and a $4 \times 4$ supercell of $\beta$-graphyne, resulting in a system with 1,072 atoms and a mismatch of only 2.8\%.

The Mulliken population analysis revealed an increase in the overlap population for $\alpha$-, $\beta$-, and $\gamma$-glaphynes, with average shifts of approximately 0.87, 10.88, and 0.96~m$|e|$, respectively. These shifts are primarily attributed to modifications in the distribution of the carbon 2p$_z$ orbitals when compared to their pristine graphyne counterparts. Notably, the most significant variation occurs in $\beta$-glaphyne, where the substantial change in C-C bond lengths within the acetylenic linkages indicates a corresponding alteration in hybridization. This result is consistent with the enhanced electronic coupling observed in the $\beta$ configuration.

From an energetic standpoint, Table~\ref{table:structural_stability} indicates that among the pristine structures, $\gamma$-glaphyne exhibits the lowest cohesive energy (see Tab. \ref{table:structural_stability}), which is defined as
\begin{equation}
    E_{\text{coh}} = \frac{1}{n_{\text{atoms}}} \left( E_{\text{Total}} -\sum_{i=1}^{n_{\text{atoms}}} E_{i}^{\text{atom}} \right),
\end{equation}
where $E_{i}^{\text{atom}}$ corresponds to the energy of each isolated atomic species, $E_{\text{Total}}$ is the total energy of the optimized structure, and $n_{\text{atoms}}$ is the total number of atoms. For the glaphynes, although all three configurations display comparable cohesive energies, their values are slightly lower (more negative) than those of the isolated graphynes. This subtle reduction may reflect enhanced structural stability, possibly promoted by the interaction between the silica overlayer and the underlying graphyne substrate \cite{SiO}.

\begin{table*}[htb]
    \centering
    \caption{Structural and energetic properties obtained in this work. Values in parentheses correspond to reference data from the literature. Listed are: number of atoms per unit cell ($n_{\text{atoms}}$), lattice parameter $a = b$ (\r{A}), angle $\gamma$ ($^\circ$), band gap energy $E_{\text{gap}}$ (eV), and cohesive energy $E_{\text{coh}}$ (eV/atom).}
    \begin{tabular}{lccccc}
        \hline
        Material & $n_{\text{atoms}}$ & $a = b$ (\r{A}) & $\gamma$ (°) & $E_{\text{gap}}$ (eV) & $E_{\text{coh}}$ (eV/atom) \\
        \hline
        $\alpha$-graphyne  & 8   & 7.02 (6.98 \cite{Kim2012,Chen2013})  & 119.9 (120 \cite{Kim2012,Chen2013}) & 0.0 (0.0 \cite{Kim2012,Chen2013})  & $-8.13$ \\
        $\beta$-graphyne   & 18  & 9.57 (9.50 \cite{Kim2012,Chen2013})  & 119.9 (120 \cite{Kim2012,Chen2013}) & 0.0 (0.0 \cite{Kim2012,Chen2013})  & $-8.20$ \\
        $\gamma$-graphyne  & 12  & 6.93 (6.88 \cite{Kim2012,Enyashin2011})  & 119.9 (120 \cite{Kim2012,Enyashin2011}) & 1.47 (1.32 \cite{Kim2012,Enyashin2011})  & $-8.41$ \\
        $\alpha$-glaphyne  & 264 & 21.19 & 120.01 & 0.01 & $-8.74$ \\
        $\beta$-glaphyne   & 66  & 10.27 & 119.98 & 1.16 & $-8.65$ \\
        $\gamma$-glaphyne  & 300 & 21.02 & 119.99 & 1.51 & $-8.77$ \\
        \hline
    \end{tabular}
    \label{table:structural_stability}
\end{table*}

\begin{figure}[t!]
\centering
\includegraphics[width=0.75\linewidth, keepaspectratio]{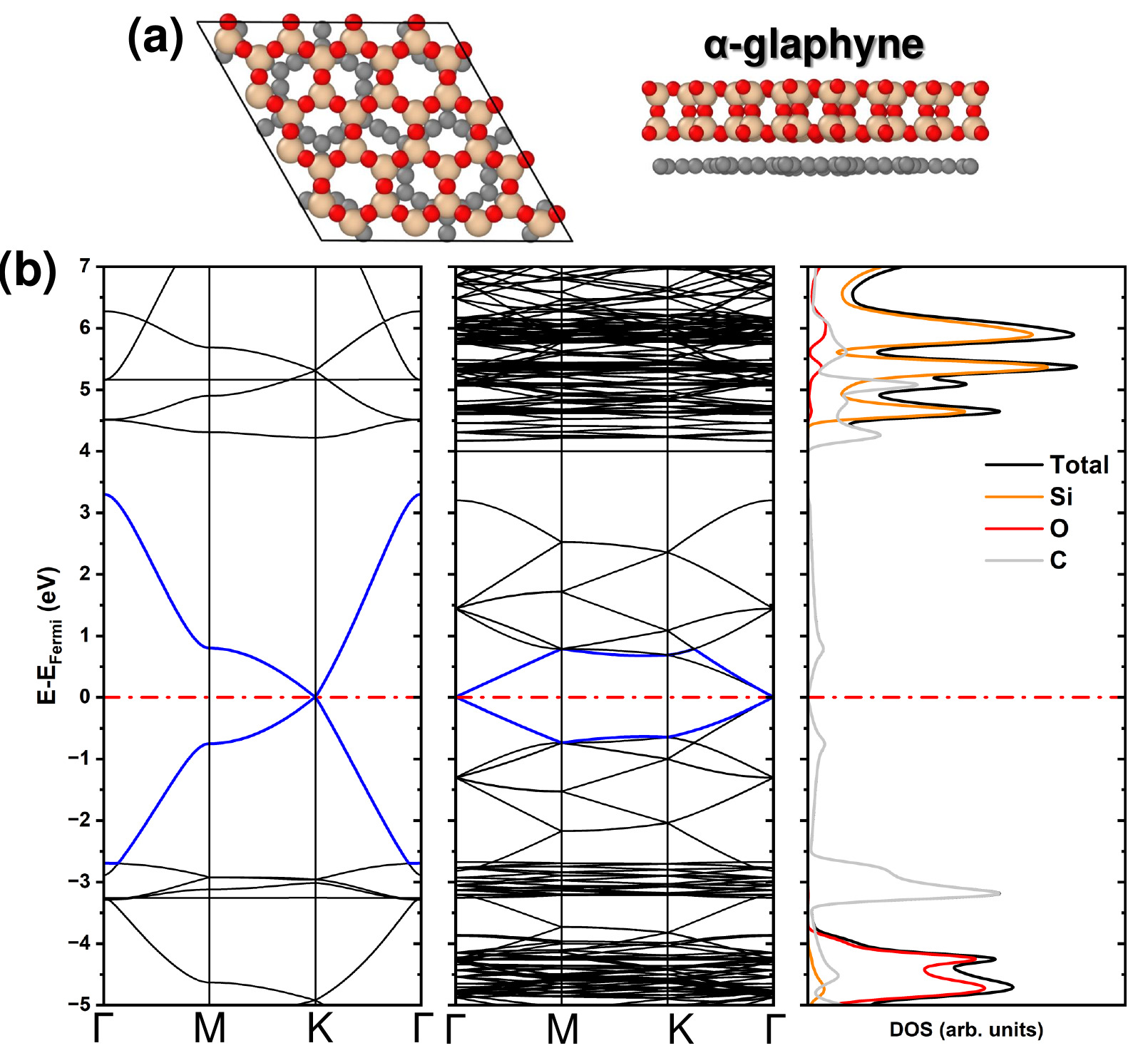}
\caption{(a) Top and side views of the $\alpha$-glaphyne unit cell. (b) Electronic band structure of $\alpha$-graphyne and $\alpha$-glaphyne, with the corresponding density of states of $\alpha$-glaphyne.}
\label{fig:fig2_bands-alfa-glaphyne}
\end{figure}

\begin{figure}[t!]
\centering
\includegraphics[width=0.75\linewidth, keepaspectratio]{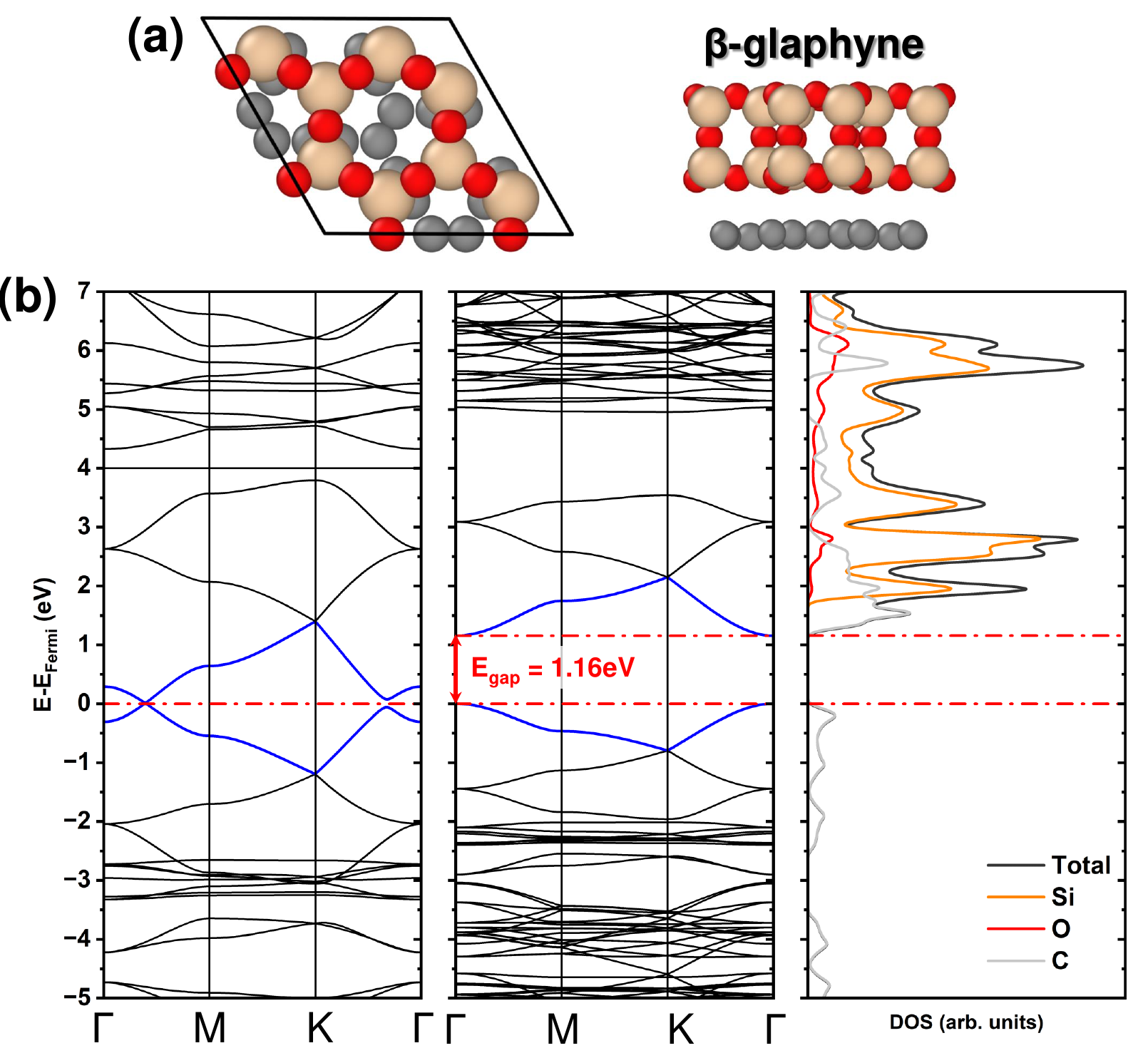}
\caption{(a) Top and side views of the $\beta$-glaphyne unit cell. (b) Electronic band structure of $\beta$-graphyne and $\beta$-glaphyne, accompanied by the density of states of $\beta$-glaphyne.}
\label{fig:fig3_bands-beta-glaphyne}
\end{figure}

\begin{figure}[t!]
\centering
\includegraphics[width=0.75\linewidth, keepaspectratio]{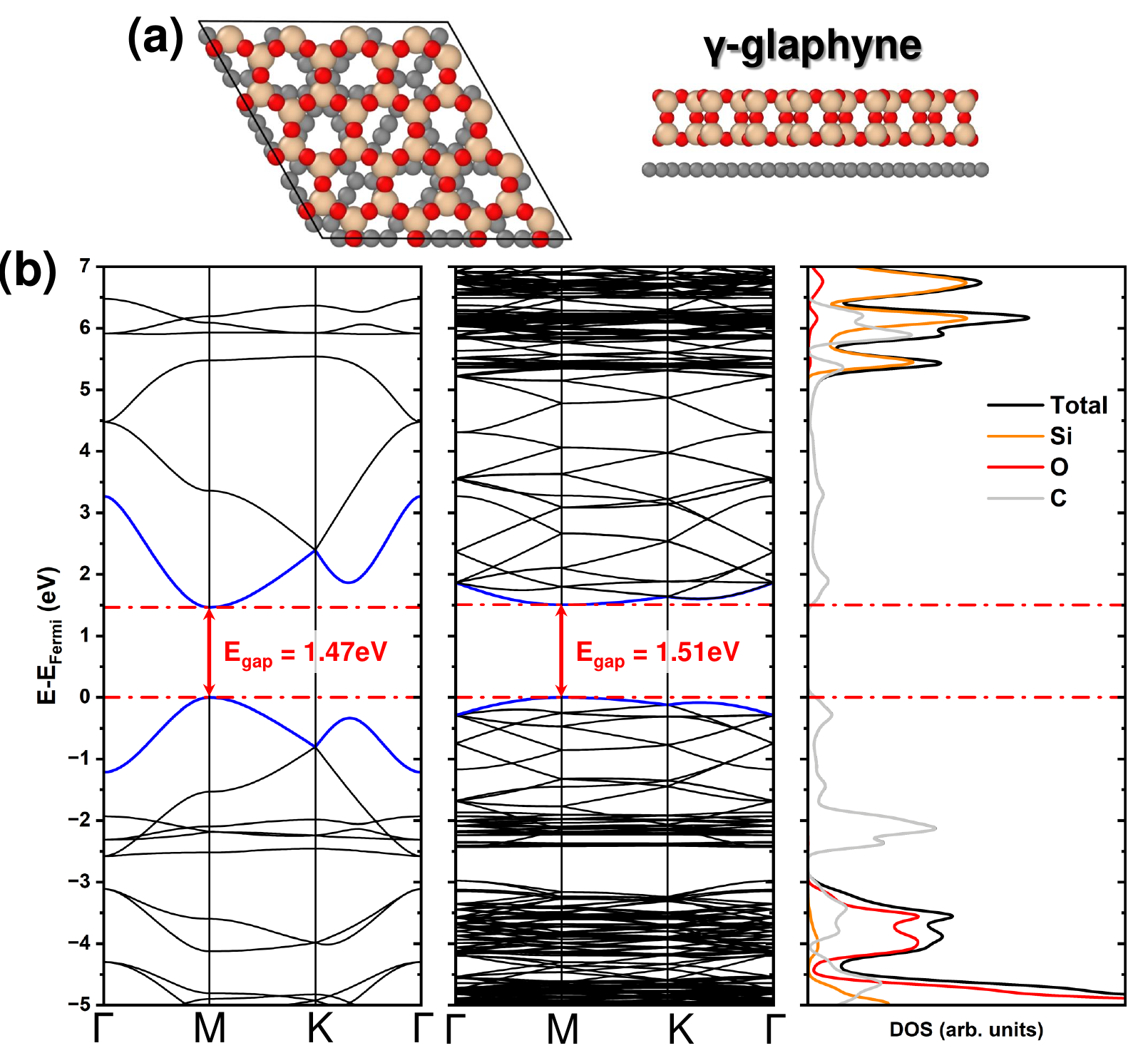}
\caption{(a) Top and side views of the $\gamma$-glaphyne unit cell. (b) Electronic band structure of $\gamma$-graphyne and $\gamma$-glaphyne, with the corresponding density of states of $\gamma$-glaphyne.}
\label{fig:fig4_bands-gamma-glaphyne}
\end{figure}

An important aspect to assess in these novel structures is their electronic behavior. In Figures~\ref{fig:fig2_bands-alfa-glaphyne}, \ref{fig:fig3_bands-beta-glaphyne}, and \ref{fig:fig4_bands-gamma-glaphyne}, we present the electronic band structures and the corresponding density of states (DOS) for $\alpha$-, $\beta$-, and $\gamma$-glaphyne, respectively.

The $\alpha$-graphyne presents a zero band gap, featuring a Dirac cone at the K-point (see Fig. \ref{fig:fig2_bands-alfa-glaphyne}(b)). In contrast, $\alpha$-glaphyne retains a Dirac cone structure, now centered at the $\Gamma$-point, while maintaining an almost zero band gap ($\sim$10 meV). The DOS analysis indicates that, although the presence of the SiO$_2$ layer alters the band structure, the states near the Fermi level remain predominantly composed of the carbon $2p$ orbitals. This suggests that the presence of the silica monolayer induces only minor perturbations in the electronic structure of $\alpha$-graphyne. Nevertheless, silicon contributes significantly to the conduction band through its $3d$ orbitals, while the valence band remains mainly composed of states from carbon and oxygen atoms.

In the case of $\beta$-graphyne, figure \ref{fig:fig3_bands-beta-glaphyne}(b) shows that the system exhibits a Dirac cone along the $\Gamma$-M path, maintaining a zero band gap, consistent with previously reported theoretical results \cite{Kim2012, Chen2013}. In contrast, $\beta$-glaphyne exhibits a direct band gap at the $\Gamma$-point, with a value of 1.16 eV, while preserving the overall band structure, albeit with slight band flattening, which leads to reduced electron mobility. DOS calculations indicate that the primary contributors to the band gap are the C $2p$ orbitals in the valence band. Nevertheless, the conduction band is mainly composed of C $2p$ orbitals, with additional contributions from Si $3d$ orbitals. Oxygen contributes only to deeper energy levels within the conduction band. Despite substantial modifications in the electronic structure, the system retains the symmetry of its pristine configuration. The band gap opening may result from strain induced in the C-C bonds due to the lattice mismatch. However, although this strain did not lead to structural deformation and the overall symmetry was preserved, the single C-C bonds experienced more significant stretching. Additionally, interlayer interactions play a crucial role, further contributing to the observed variation in the band gap.

Finally, for the $\gamma$-graphyne, a direct band gap of 1.47 eV is observed at the M-point, in good agreement with previous studies \cite{Kim2012, Enyashin2011}. As shown in Figure \ref{fig:fig4_bands-gamma-glaphyne}(b), the presence of SiO$_2$ induces a slight modification in the band structure, characterized by a band flattening and an increase in the band gap to 1.51 eV, still located at the M-point. DOS calculations reveal a composition similar to that of $\alpha$-glaphyne, with the $2p$ orbitals of carbon atoms dominating the states near the Fermi level. Contributions from silicon and oxygen appear only in deeper energy levels, suggesting that their influence remains confined to lower-lying electronic states.

\begin{figure}[t!]
\centering
\includegraphics[width=0.5\linewidth, keepaspectratio]{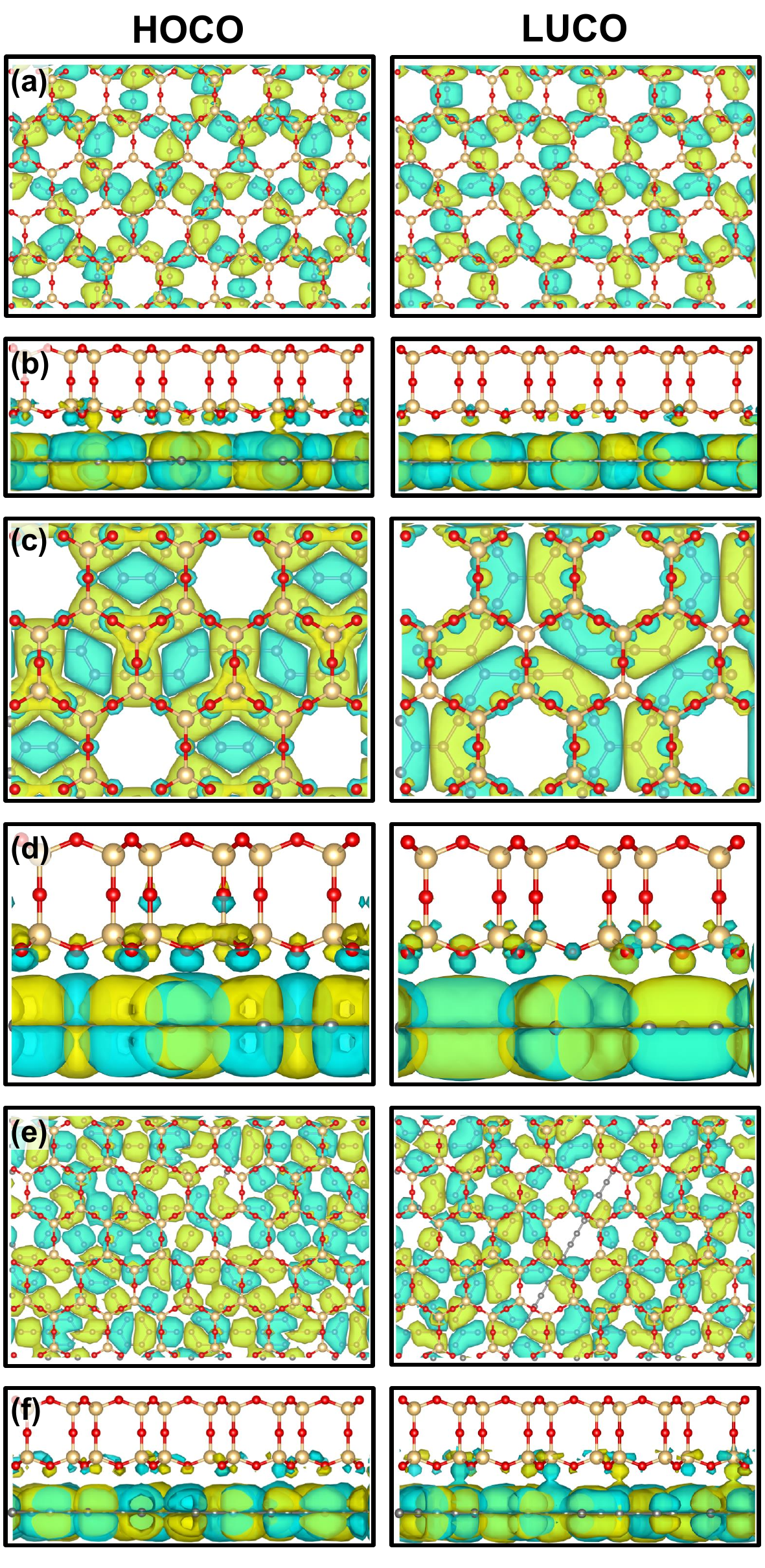}
\caption{Highest Occupied Crystalline Orbital (HOCO) and Lowest Unoccupied Crystalline Orbital (LUCO) isosurfaces for glaphynes. Panels (a) and (b) show the top and side views of $\alpha$-glaphyne, (c) and (d) correspond to $\beta$-glaphyne, and (e) and (f) to $\gamma$-glaphyne, respectively. Blue and yellow isosurfaces represent positive and negative charge densities.}
\label{fig:fig5_hoco-luco-topview}
\end{figure}

Another crucial property to examine is the spatial distribution of the frontier orbitals, namely the Highest Occupied Crystalline Orbital (HOCO) and the Lowest Unoccupied Crystalline Orbital (LUCO), which are presented in Figure~\ref{fig:fig5_hoco-luco-topview}. For $\alpha$-glaphyne (Figures~\ref{fig:fig5_hoco-luco-topview}(a) and \ref{fig:fig5_hoco-luco-topview}(b)) and $\gamma$-glaphyne (Figures~\ref{fig:fig5_hoco-luco-topview}(e) and \ref{fig:fig5_hoco-luco-topview}(f)), both HOCO and LUCO exhibit a relatively delocalized character, with HOCO showing slightly greater localization. In contrast, $\beta$-glaphyne displays highly localized and well-defined frontier orbitals, which may favor improved electron transport. This behavior can be attributed to its stacking configuration, in which the pores of the SiO$_2$ and $\beta$-graphyne layers are periodically aligned. Additionally, Figures~\ref{fig:fig5_hoco-luco-topview}(c) and \ref{fig:fig5_hoco-luco-topview}(d) reveal a slight charge redistribution from the graphyne layer to the SiO$_2$, further reinforcing the system’s structural integrity and indicating that the electronic characteristics of the pristine graphynes are largely preserved.

\begin{figure}[t!]
\centering
\includegraphics[width=0.8\linewidth, keepaspectratio]{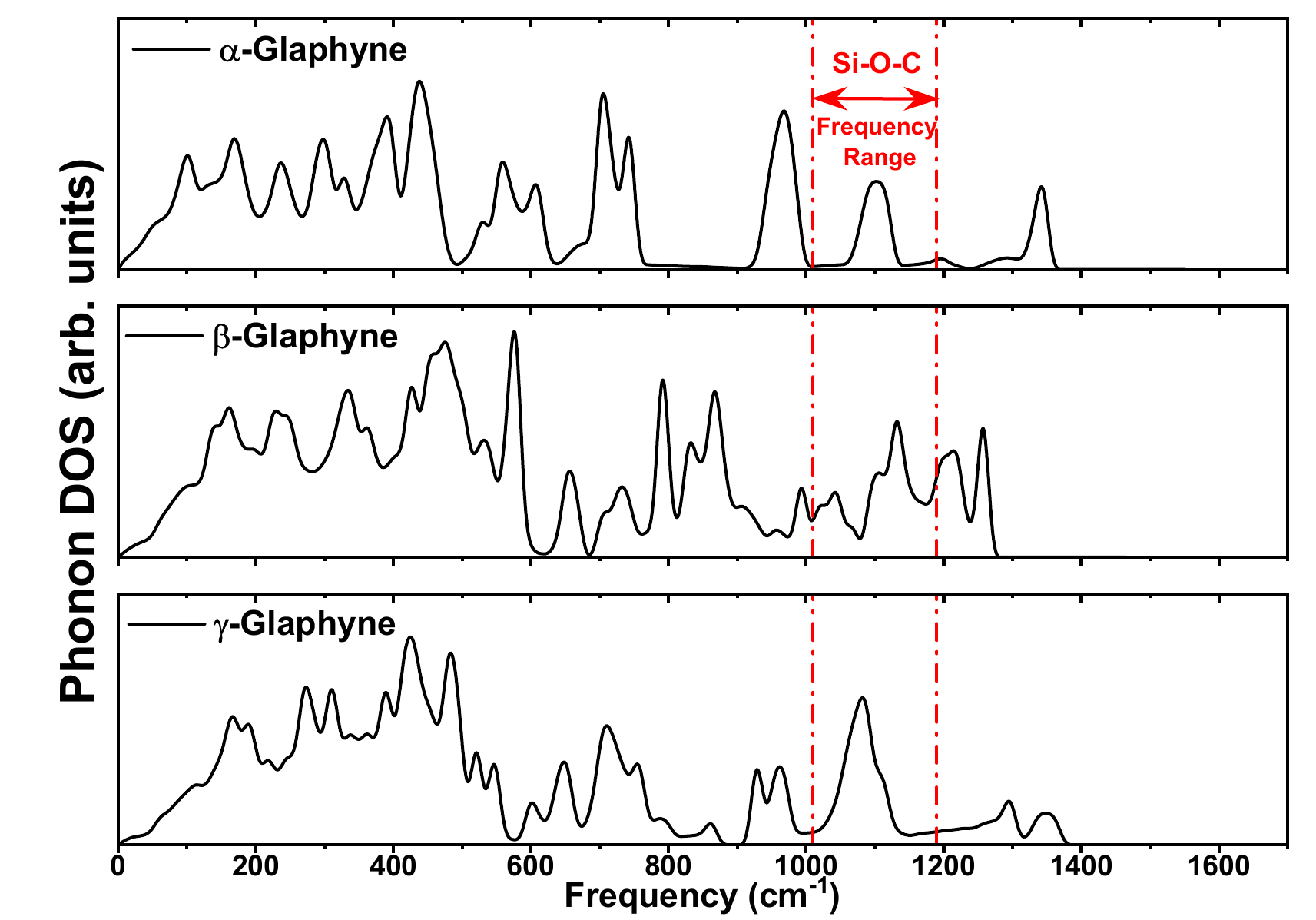}
\caption{Phonon density of states of $\alpha$-, $\beta$-, and $\gamma$-glaphyne. The red dotted region highlights the Si-O-C frequency range.}
\label{fig:fig7_phononDOS}
\end{figure}

In order to unveil and compare the behavior of glaphynes with that of glaphenes, Figure~\ref{fig:fig7_phononDOS} presents the phonon density of states (DOS) of $\alpha$-, $\beta$-, and $\gamma$-glaphyne. This analysis aims to understand the vibrational behavior better and, in particular, to verify the presence of Si-O-C bonding features, similar to those reported by Iyengar \textit{et al.}~\cite{Iyengar2025}. Interestingly, although the three glaphyne structures share common characteristics—such as the presence of both low- and high-frequency vibrational modes—distinct differences emerge in the distribution and intensity of the phonon peaks. 

The $\alpha$-glaphyne exhibits well-defined and intense peaks across several frequency regions, with a notable concentration in the high-frequency range (900-1200~cm$^{-1}$). In contrast, $\beta$-glaphyne shows a more dispersed peak distribution, with generally lower intensities, yet retains a well-marked high-frequency region with a slightly different spectral profile. Meanwhile, $\gamma$-glaphyne presents behavior similar to $\beta$-glaphyne in the low-frequency range, with less pronounced peaks than $\alpha$-glaphyne, while still displaying distinct vibrational features across the spectrum. The similarities between $\alpha$- and $\gamma$-glaphynes may be attributed to their larger unit cells and the periodic alignment of the pores in the stacking configuration.

A key spectral region in the phonon DOS lies between approximately 950 and 1150~cm$^{-1}$, which is primarily associated with Si-O-C bonding. Vibrational modes in this range are typically attributed to Si-O and C-O bond stretching, as well as coupled modes involving all three elements. In $\alpha$-glaphyne, pronounced peaks appear near 980 and 1080~cm$^{-1}$, indicating strong Si-O-C bonding. The intensity and position of these peaks serve as signatures of the formation and structural stability of such bonds, resembling those observed in glaphene.

Notably, while the phonon DOS profiles of glaphynes and glaphene share some similarities, the G-band and the D'-like shoulder appear shifted toward higher frequencies. This displacement may be attributed to the parametrization adopted in this study, which, as with most parametrizations, is tuned for general-purpose simulations and may introduce deviations in specific vibrational modes. When considered alongside the Mulliken population analysis, the phonon data strongly suggest that $\beta$-glaphyne exhibits the most pronounced Si-O-C bonding character. This is evidenced by a distinct peak at 1132~cm$^{-1}$ and the emergence of a secondary shoulder in the same region, reinforcing the relevance of the Si-O-C bond formation to the structural and electronic properties of the system. For comparison, in the case of glaphene, a characteristic frequency associated with the same interaction is observed around 1050~cm$^{-1}$~\cite{Iyengar2025}.

\section{Conclusion and perspectives}  

Using the DFTB methodology, we proposed a new class of materials referred to as glaphynes, similar to glaphenes. The comparative analysis of $\alpha$-, $\beta$-, and $\gamma$-glaphyne reveals distinct electronic modifications induced by the growth of SiO$_2$, with potential implications for functional material applications. While $\alpha$-glaphyne retains a nearly zero band gap and exhibits a redistribution of electronic states near the Fermi level, $\beta$-glaphyne undergoes a significant band gap opening of 1.16~eV, accompanied by band flattening that may reduce electron mobility. In contrast, $\gamma$-glaphyne maintains a direct band gap, which increases slightly to 1.51~eV upon SiO$_2$ incorporation, suggesting non-pronounced quantum confinement effects.

These findings demonstrate that structural modifications achieved through heterostructure engineering enable the precise tuning of electronic properties, thereby influencing charge transport and band characteristics. The band gap variations and orbital rearrangements observed in these systems highlight the potential of glaphyne-based heterostructures for use in nanoelectronic devices, semiconducting applications, and surface-mediated catalysis, where control over band alignment and electronic coupling is essential.

In summary, we have observed that in the case of glaphynes, the electronic proximity effect can indeed open the electronic band gap, but not for all cases, even with the formation of Si-O-C bonds. These results highlight the potential of using the electronic proximity effect to create new engineering band gap heterostructures.

\begin{acknowledgement}
Guilherme S. L. Fabris acknowledges the São Paulo Research Foundation (FAPESP) fellowship (process number $\#$2024/03413-9). Raphael B. de Oliveira thanks the National Council for Scientific and Technological Development (CNPq) (process numbers 151043/2024-8 and 200257/2025-0). Marcelo L. Pereira Junior acknowledges financial support from FAPDF (grant 00193-00001807/2023-16), CNPq (grants 444921/2024-9 and 308222/2025-3), and CAPES (grant 88887.005164/2024-00). Douglas S. Galvão acknowledges the Center for Computing in Engineering and Sciences at Unicamp for financial support through the FAPESP/CEPID Grant (process number $\#$2013/08293-7). We thank the Coaraci Supercomputer Center for computer time (process number $\#$2019/17874-0).
\end{acknowledgement}

%

\bibliography{bibliography.bib}

\end{document}